\documentclass[prl,comment,twocolumn]{revtex4}
\usepackage{graphicx}
\usepackage{dcolumn}
\usepackage{bm}
\usepackage{amsmath}
\usepackage{amssymb}
\usepackage{mathrsfs}
\usepackage{amsfonts}
\usepackage{url}
 
\def\ket#1{\mathinner{|{#1}\rangle}}

\begin{document} 

\title{Comment on "Superradiant Phase Transitions and the Standard Description of Circuit QED" }

\author{Cristiano Ciuti}
\author{Pierre Nataf}
\affiliation{Laboratoire Mat\'eriaux et Ph\'enom\`enes Quantiques,
Universit\'e Paris Diderot-Paris 7 and CNRS, \\ B\^atiment Condorcet, 10 rue
Alice Domon et L\'eonie Duquet, 75205 Paris Cedex 13, France}

\maketitle
In a recent letter\cite{Viehmann}, Viehmann, von Delft and Marquardt claim that a superradiant  phase transition\cite{Brandes} for a chain of Cooper pair boxes {\it capacitively} coupled to a transmission line resonator is prevented by the Thomas-Reiche-Kuhn (TRK)  sum rule for the {\it electron} dipole operator. The authors consider a closed electron box model for the  artificial atoms coupled to the quantized vector potential of a cavity photon mode. Since the authors are aware that the standard model of Cooper pair boxes\cite{Blais} predicts such transition\cite{Nataf},  they conclude that such standard Hamiltonian description in circuit quantum electrodynamics (QED) must have a problem in the limit of a large number $N$ of Cooper pair boxes capacitively coupled to a resonator (the phase transition is expected to occur for $N \gg 1$). We believe that the main claim in Ref.  \cite{Viehmann} is unproven as argued below. 

(i) While the superradiant phase transition for Cooper pair boxes involves the superconducting degrees of freedom and dressed fields in the circuit, the authors do not consider at all any link between the TRK sum rule for the electron dipole operator matrix elements and the number of Cooper pairs or the superconducting phase difference across the Josephson junctions. By contrast, in the cavity QED case with real atoms\cite{Nataf}, the link between the TRK electron sum rule and the superradiant phase transition is direct, because the transition affects directly the electron dipole operator associated to the atomic two-level systems. Moreover, it is unfortunate that Viehmann et al.\cite{Viehmann} do not show at all how the TRK sum rule for the electron dipole operator modifies the standard quantization of circuit QED Hamiltonians.

(ii) In their section 'Microscopic description of Circuit QED' (page 3 of Ref. \cite{Viehmann}), Viehmann {\it et al.}  describe via the Hamiltonian ${\mathcal H}_{\rm mic}$ each artificial atom as a closed box with a fixed number $n_k$ of electrons ($k = 1,2,...,N$ is the artificial atom index), hence they take the same kind of Hamiltonian for the cavity QED case as in the first part of their letter\cite{Viehmann} and as in the first part of Ref. \cite{Nataf}.  It is surprising that the  authors\cite{Viehmann} in their microscopic model do not consider the role of the gate chemical potential, controlling the population of Cooper pairs in the superconducting box. In fact, for a proper gate voltage and for a capacitance energy $E_C$ much larger than the Josephson energy $E_J$, the ground state $\ket{g}$ and first excited state $\ket{e}$ of a Cooper pair box are $\ket{g} \propto (\ket{n} + \ket{n+1})$, $\ket{e} \propto( \ket{n} - \ket{n+1})$, where $n$ is an  integer number of Cooper pairs which have Josephson tunneled into the box: hence the number of Cooper pairs
is not fixed in such states. 

(iii) The generalization of the cavity QED no-go theorem to the multilevel atomic case with an arbitrary spectrum presented in Ref. \cite{Viehmann} is based on the assumption that in the thermodynamic limit it is always possible to neglect the  transitions between excited atomic levels to determine the existence of the superradiant phase transition. We show here a concrete counterexample showing that such key assumption in Ref. \cite{Viehmann} can be violated. Let us consider a generalized Dicke model with $N$  3-level system $\{|0_k\rangle, |1_k\rangle, |2_k\rangle\}$ (for $k=1...N$) with the transitions $|0_k\rangle \rightarrow |1_k\rangle$ and $|1_k\rangle \rightarrow |2_k\rangle$ coupled to the same photon mode via the coupling $\lambda_{01}$ and $\lambda_{12}$ respectively and no optical coupling between $|0_k\rangle$ and $ |2_k\rangle$ ($\lambda_{02}=0$, this is the well-known ladder configuration). If $\lambda_{01} =0 $, the system can be solved with known methods\cite{Brandes} since the two levels $|1_k\rangle$ and $|2_k\rangle$ coupled to the same photon mode form a Dicke system:  for $\lambda_{12}$ above a critical coupling, a superradiant phase transition can occur with a ground state involving only the states $|1_k\rangle$ and $|2_k\rangle$ and {\it not} $|0_k\rangle$, while in the normal phase only the $|0_k\rangle$ state is 
populated.  This is a first order phase transition and so the population
of excited atomic levels in the ground state jumps discontinuously to a macroscopic number.  Such a
jump invalidates a perturbative expansion in terms of number of excited
atoms as done in Ref. \cite{Viehmann}. At least for $\lambda_{01} \ll  \lambda_{12}$, a first-order transition can occur in a such a counterexample model, with the dominant weight in the superradiant ground states carried by the excited $\vert1_k\rangle \to |2_k\rangle$ transition. Note  that  the treatment in Ref. \cite{Viehmann} considers only the role of the squared vector potential ${\bf \hat{A}}^2$ term for transitions from the ground level $\vert 0 \rangle$; since the TRK sum rule implies different, less stringent inequalities for the strength of the
excited state transitions, the reasoning by Viehmann et al. does not cover this counterexample. Using the language of phase transitions, for the multilevel case Viehmann et al.\cite{Viehmann} have implicitly considered only the case of second-order phase transitions (as in the standard case with two-level systems\cite{Brandes}), while with multilevel systems superradiant transitions of the first-order with a more abrupt change of the ground state properties need to be considered, as explicitly shown by our $3$-level counterexample for the case of a single photon mode or by other closely related systems \cite{Brandes3}. The issue of multilevel superradiant phase transitions is therefore open even in atomic cavity QED and needs to be explored further in the future.

In conclusion,  while we agree with the authors of Ref. \cite{Viehmann} that we should encourage experiments to test the validity of the standard Hamiltonian of circuit QED,  the theoretical treatment in Ref. \cite{Viehmann} misses to show how the bare electron dipole TRK sum rule affects the quantization of circuit QED Hamiltonians. Moreover, the {\it generalized} no-go theorem for multilevel systems in Ref. [1] is based on an assumption, which is not general as shown by an explicit 3-level counterexample.



\end{document}